\documentclass{IEEEtran}
\usepackage{cite}
\usepackage{amsmath,amssymb,amsfonts}
\usepackage{algorithmic}
\usepackage{graphicx}
\usepackage{textcomp}
\def\BibTeX{{\rm B\kern-.05em{\sc i\kern-.025em b}\kern-.08em
    T\kern-.1667em\lower.7ex\hbox{E}\kern-.125emX}}
\usepackage{color}
\usepackage{bm}
\usepackage{algorithm}
\usepackage{algorithmic}

\ifCLASSOPTIONcompsoc
    \usepackage[caption=false, font=normalsize, labelfont=sf, textfont=sf]{subfig}
\else
\usepackage[caption=false, font=footnotesize]{subfig}
\fi

\begin{document}
\title{Radar Imaging by Sparse Optimization Incorporating MRF Clustering Prior}
\author{Shiyong Li, \IEEEmembership{Member, IEEE}, Moeness Amin, \IEEEmembership{Fellow, IEEE}, 
Guoqiang Zhao, 
and Houjun Sun
\thanks{Manuscript received , 2018. This work was performed while Dr. S. Li was a Visiting Research Scholar in the Center for Advanced Communications, Villanova University. The work of Shiyong Li, Guoqiang Zhao, and Houjun Sun was supported by the National Natural Science Foundation of China under Grant 61771049.}
\thanks{S. Li, G. Zhao and H. Sun are with the Beijing Key Laboratory of Millimeter Wave and Terahertz Technology, Beijing Institute of Technology, Beijing 100081, China. (e-mail: lisy\_98@bit.edu.cn).}
\thanks{M. Amin is with the Center for Advanced Communications, Villanova University, Villanova, PA 19085 USA (e-mail: moeness.amin@villanova.edu).}
}

\maketitle

\begin{abstract}
Recent progress in compressive sensing states the importance of exploiting intrinsic structures in sparse signal reconstruction. In this letter, we propose a Markov random field (MRF) prior in conjunction with fast iterative shrinkage-thresholding algorithm (FISTA) for image reconstruction. The MRF prior is used to represent the support of sparse signals with clustered nonzero coefficients. 
The proposed approach is applied to the inverse synthetic aperture radar (ISAR) imaging problem. 
Simulations and experimental results are provided to demonstrate the performance advantages of this approach in comparison with the standard FISTA and existing MRF-based methods. 
\end{abstract}


\section{Introduction}
\label{sec:introduction}
\IEEEPARstart{T}{he} theory of compressive sensing (CS) \cite{b13,b14} has generated great interest in radar imaging \cite{baraniuk_cs_radar,Herman_2009,Potter_2010,Amin_2010book,Yoon_2010,Moreira_2013,Amin_2014book,Cetin_2014}. Compressive sensing-based radar imaging methods can achieve better image quality than traditional Fourier transform-based techniques under image sparsity condition and with incomplete observations, as viewed from Nyquist sampling requirements.

However, in many real world scenarios, there exists other important signal structure information, beside sparsity, which can be exploited for enhanced imaging \cite{b12,b25,b28,b36}. 
For instance, most of the wavelet coefficients of a natural image are both sparse and exhibit a tree structure \cite{b12}. Clustering is a structure property in which the image assumes high values in neighboring pixels. It may be exhibited in the image canonical basis or under sparsifying basis. From sparse reconstruction perspective, this property is known as \textit{block sparse}, or \textit{clustered sparse}\cite{b25}. In essence, the support of the image spatial pixels assumes high correlations among closely separated pixels which conforms to the block sparse model. 

Image reconstruction approaches incorporating block sparse models were developed in \cite{b17,b16}. These approaches combine the $\ell_1$ and $\ell_2$ norms for feature selections, and assume knowledge of group partition information. This assumption is not mandated in the block sparse Bayesian frame work proposed in \cite{b21}, though performance could still be affected by the block size. 

Using graph theory, the radar image scene can be cast as an undirected graphical model, also referred to as Markov random field (MRF).
MRF provides an effective way for modeling spatial context dependent entities, such as image pixels and correlated features. 
It has been widely employed in a variety of areas including computer vision, image segmentation\cite{b22}, and sparse signal recovery\cite{b25,b28,b36}.

The use of MRF prior in CS based magnetic resonance imaging was recently introduced in \cite{b30}. The authors extended a constrained split augmented Lagrangian shrinkage algorithm (C-SALSA) with an MRF prior \cite{b31}, and solved the problem using the alternating direction method of multipliers (ADMM) \cite{Boyd_admm}. The analysis of the ADMM reveals the sensitivity of this type of algorithms to the penalty parameter underpinning the augmented Lagrangian where the primal and dual residuals exhibit conflicting requirements. 

In this letter, we incorporate the MRF prior with fast iterative shrinkage/thresholding algorithm (FISTA)\cite{fista_tv}. Unlike ADMM, the penalty parameter of FISTA is used to balance the twin objectives of minimizing both the $l_2$ error and signal sparsity, without significantly affecting the algorithm convergence.
FISTA also preserves the computational simplicity of ISTA family, and achieves a quadratic convergence rate.
To demonstrate performance, we apply the proposed MRF-based FISTA algorithms to inverse synthetic aperture radar (ISAR) imaging, in which scatterers are typically located contiguously, satisfying the clustering property. 

The remainder of this letter is as follows. Section II presents the formulation of the MRF-based FISTA. In Section III, we apply the MRF-FISTA to ISAR imaging. Numerical simulations and experimental results are shown in Section IV. Finally, concluding remarks are presented in Section V.

\section{MRF-based FISTA}
In compressive sensing, we aim to recover a signal $\mathbf{x}\in\mathbb{C}^{\mathnormal{N}}$ from $M\leq N$ noisy linear measurements $ \mathbf{y}\in\mathbb{C}^{\mathnormal{M}} $,
\begin{equation} \label{eq1}
\mathbf{y}=\mathbf{\Phi x}+\mathbf{\epsilon}
\end{equation}
where $\mathbf{\Phi}\in \mathbb{C}^{\mathnormal{M}\times \mathnormal{N}}$ denotes a known sensing matrix, and $\mathbf{\epsilon} \in \mathbb{C}^\mathnormal{M}$ is white Gaussian noise. 
In the following analysis, we assume $\mathbf{x}$ is sparse in its canonical basis for convenience. If not, we can apply a sparsifying transform $\mathbf{\Psi}$, such as $\mathbf{x}\!=\!\mathbf{\Psi}\bm{\alpha} $, where $\bm{\alpha}$ is a sparse vector.

The estimation of $\mathbf{x}$ from $\mathbf{y}$ in (\ref{eq1}) is an ill-posed linear inverse problem, since the sensing matrix $\mathbf{\Phi}$ is singular or ill-conditioned. However, if $\mathbf{x}$ is sparse or has sparse coefficients in some basis $\mathbf{\Psi}$, then exact solution of $\mathbf{x}$ can be obtained by solving the following relaxed convex optimization problem,
\begin{equation} \label{eq2}
\mathbf{\hat{x}} = \rm{arg}\mathop{\rm{min}}\limits_{\mathbf{x}}\left\{\frac{1}{2} \Vert{\mathbf{y}-\mathbf{\Phi x}} \Vert_2 ^2 + \lambda \Vert{\mathbf{x}}\Vert_1  \right\}
\end{equation}
where $\Vert\mathbf{x}\Vert_1 = \sum_i\vert{x}_i\vert $ denotes the $\ell_1$ norm of $\mathbf{x}$, and $\Vert\mathbf{y}\Vert_2 = ( \sum_i\vert{y}_i\vert ^2 ) ^{1/2}$ represents the $\ell_2$ norm of $\mathbf{y}$. 

\begin{figure}[!t]
	\centering
	\includegraphics[width=2.6in]{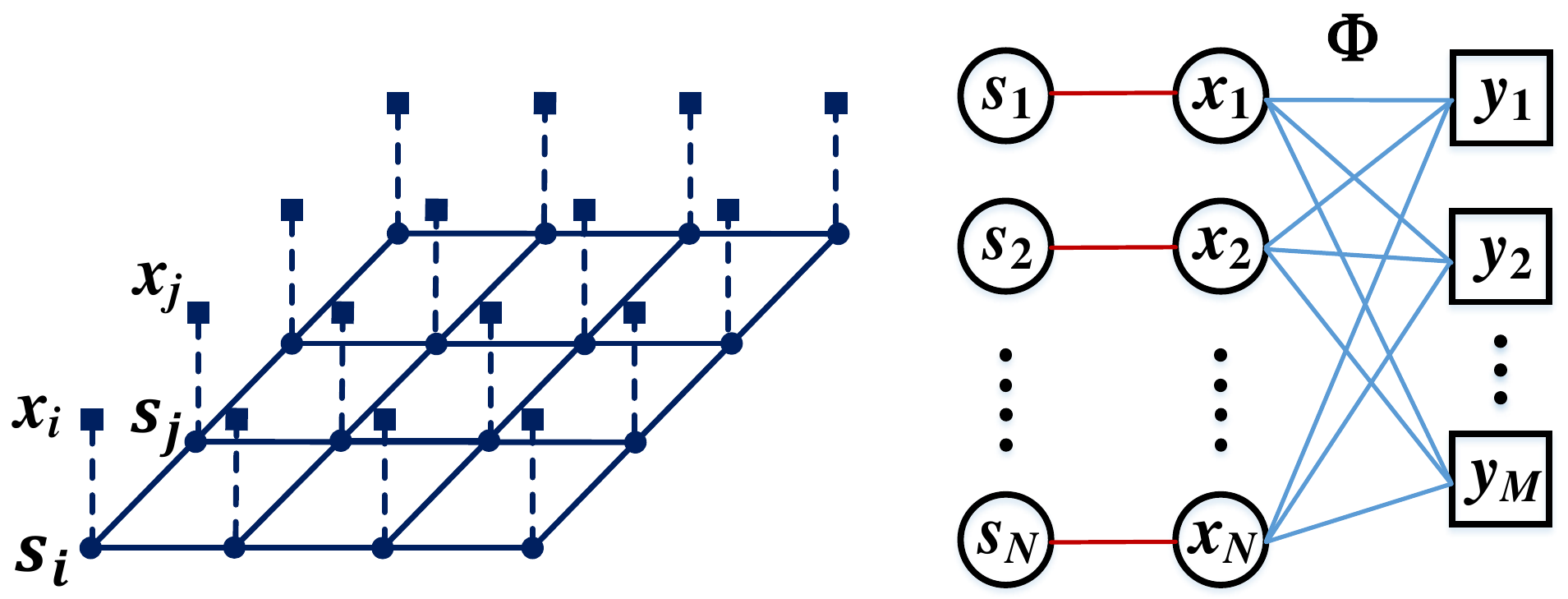}
	\caption{A graphical model of variables, operators and their relations.}
\label{fig1}
\end{figure}

To invoke the clustering property, we assume the probability distribution of the prior structural information as $p(\mathbf{s})$, with $\mathbf{s}$ representing the signal support. The relationship among the signal $\mathbf{x}$, the signal support $\mathbf{s}$, and the measurement $\mathbf{y}$ can be illustrated by an undirected graphical model, shown in Fig. \ref{fig1}. Independence between two variables is displayed as a lack of connection between their corresponding vertices in the graph; conversely,  dependent variables should correspond to connected vertices. Further, the absence of a direct link between two variables reflects an interaction that is conditional. In this regard, and for the underlying problem, given signal $\mathbf{x}$, the signal support $\mathbf{s}$ and the measurement $\mathbf{y}$ are independent. Accordingly, the maximum \textit{a posterior} (MAP)  estimation of the signal support $\mathbf{s}$ becomes
\begin{align} \label{esti_s}
\mathbf{\hat{s}}&=\arg\max\limits_{\mathbf{s}}p(\mathbf{s}|\mathbf{x},\mathbf{y}) =\arg\max\limits_{\mathbf{s}}p(\mathbf{s}|\mathbf{x}) \nonumber \\
&=\arg\max\limits_{\mathbf{s}}p(\mathbf{x}|\mathbf{s})p(\mathbf{s}).
\end{align}

The aforementioned undirected graph, referred to as Markov random field, can be described by an Ising model \cite{b30}, 
\begin{equation} \label{p_s}
p(\mathbf{s};\alpha,\beta)\!=\!\frac{1}{Z}e^{-\frac{1}{T} \big[\sum_{i\in{V}}V_1(s_i)\!+\!\sum_{(i,j)\in{E}}{V_2(s_i,s_j)}\big]}
\end{equation}
where each vertex has two possible states $\mathbf{s}\in \{0,1\}^N$, $Z$ and $T$ are constants. The content in the ``[]'' is the \textit{energy function}, which is a sum of \textit{potentials} over the single-site cliques $V$ and the pair-site cliques $E$, defined as $V_1(s)\!=\!\pm \alpha$ corresponding to $s\!=\!0$ and $1$, respectively, and $V_2(s_i,s_j)\!=\!\pm\beta$ corresponding to $s_i\!\neq \!s_j$ and $s_i\!=\!s_j$, respectively. It is noted that a higher value of $\alpha$ enforces a sparser signal activity, and a higher value of $\beta$ implies a stronger spatial correlation. 

Substituting \eqref{p_s} into \eqref{esti_s} yields, 
\begin{align} \label{esti_s_log}
\mathbf{\hat{s}}\!=\!\arg\max\limits_{\mathbf{s}} \bigg \{\!\!-\!\!\sum_{(i,j)\in{E}}V_2(s_i,s_j)\!\!-\!\!\sum_{i\in{V}}\big[V_1(s_i)\!\!+\!\!\log(p(x_i|s_i)) \big]\bigg\}.
\end{align}
In compressive sensing, $p(\mathbf{x})$ is typically selected as Laplacian $p(\mathbf{x})=\frac{\lambda}{2} \exp\left(-\frac{\lambda}{2}\Vert {\mathbf{x}}\Vert_1\right)$ to preserve sparsity. 

Exact inference on \eqref{esti_s_log} is hard because the graph in Fig. \ref{fig1} is not a tree, so one cannot use any of the sum-product algorithms. 
However, the variables of the Ising model can be estimated using approximate inference algorithms \cite{graphical_models}. In this letter, we use the Markov Chain Monte Carlo (MCMC) sampler introduced in \cite{b22}. 
 
Upon obtaining the intermediate value of $\mathbf{s}$, the MAP estimation of $\mathbf{x}$ can proceed as,
\begin{align}\label{MAPofx}
\mathbf{\hat{x}}&\!=\!\arg\max\limits_{\mathbf{x}}p(\mathbf{x}|\mathbf{y},\mathbf{s})\!=\!\arg\max\limits_{\mathbf{x}} p(\mathbf{y},\mathbf{s}|\mathbf{x})p(\mathbf{x})  \nonumber \\
&\!=\!\arg\max\limits_{\mathbf{x}} p(\mathbf{y}|\mathbf{x})p(\mathbf{s}|\mathbf{y},\mathbf{x})p(\mathbf{x})\!=\!\arg\max\limits_{\mathbf{x}} p(\mathbf{y}|\mathbf{x})p(\mathbf{s}|\mathbf{x})p(\mathbf{x})  \nonumber \\
&\!=\!\arg\max\limits_{\mathbf{x}} p(\mathbf{y}|\mathbf{x})p(\mathbf{x}|\mathbf{s})p(\mathbf{s}) \!=\!\arg\max\limits_{\mathbf{x}}p(\mathbf{y}|\mathbf{x})p(\mathbf{x}|\mathbf{s}).
\end{align}

For white Gaussian noise with variance $\sigma^2$, and $p(\mathbf{x})$ as Laplacian, \eqref{MAPofx} can be represented by the following two steps, 
\begin{align}\label{esti_x}
\mathbf{\hat{x}'}&=\arg\max\limits_{\mathbf{x}'} \left \{-\frac{1}{2\sigma^2}\Vert {\mathbf{y}-\mathbf{\Phi x'}}\Vert_2^2-\frac{\lambda}{2}\Vert{\mathbf{x'}}\Vert_1\right\}, \nonumber \\
\mathbf{\hat{x}}&=\mathbf{\hat{x}'}\circ\mathbf{s}
\end{align}
where $\mathbf{\hat{x}}^{\prime}$ denotes the recovered signal without MRF prior,  ``$\circ$'' represents an element-to-element multiplication. Therefore, we can obtain an estimate of signal $\mathbf{x}$ from the solutions of \eqref{esti_s_log} and \eqref{esti_x}. 
Towards this end, we employ convex fast iterative shrinkage/thresholding algorithm (FISTA) \cite{fista_tv} combined with the MRF prior. FISTA has been shown to have a quadratic convergence rate, whereas the traditional ISTA family only achieves a first-order one.  

In this letter, we apply two norms, namely, the $\ell_1$ norm prior and the TV norm prior, to improve the recovered signal quality. Whereas the $\ell_1$ norm is typically used to enhance the sparsity of the signal, the TV norm is utilized as an edge preserving regularization of the signal. We consider the signal reconstruction problem,
\begin{equation}\label{eq13}
\mathbf{\hat{x}'}\!=\!\arg\min\limits_{\mathbf{x}'}\left\{\frac{1}{2}\Vert {\mathbf{y}\!\!-\!\!\mathbf{\Phi x'}}\Vert_2^2\!\!+\!\!{\lambda_1}\Vert{\mathbf{x'}}\Vert_1\!\!+\!\!\lambda_{\textrm{TV}}\Vert{\mathbf{x'}}\Vert_{\rm{TV}}\right\}
\end{equation}
where the parameter $\sigma^2$ in \eqref{esti_x} is absorbed into $\lambda_1$ and $\lambda_{\textrm{TV}}$.

Based on approximate Taylor expansion \cite{fista_tv}, \eqref{eq13} can be transformed to,
\begin{equation}\label{eq16}
\mathbf{\hat{x}'}\!=\!\arg\min\limits_{\mathbf{x}'} \left \{\frac{L}{2}\Vert {\mathbf{x^{\prime}}\!\!-\!\!\mathbf{v}_k}\Vert_2^2\!\!+\!\!{\lambda_1}\Vert{\mathbf{x'}}\Vert_1\!\!+\!\!\lambda_{\textrm{TV}}\Vert{\mathbf{x'}}\Vert_{\rm{TV}}\right\}
\end{equation}
where $\mathbf{v}_{k} = \mathbf{x^{\prime}}_{k-1}-\frac{1}{L}\nabla f(\mathbf{x^{\prime}}_{k-1})$, $\mathbf{x}^{\prime}_{k} $ represents the estimated value of $\mathbf{x}^{\prime}$ at the $k\textrm{th}$ iteration,  $\nabla f(\mathbf{x^{\prime}})=\mathbf{\Phi}^{H}(\mathbf{\Phi x^{\prime}}-\mathbf{y})$ representing the gradient of $f(\mathbf{x^{\prime}})$, $L$ is the Lipschitz constant of $\nabla f$, and $f(\mathbf{x^{\prime}})=\frac{1}{2}\Vert {\mathbf{y}-\mathbf{\Phi x'}}\Vert_2^2$.

To solve for $\mathbf{x}^{\prime}$, we apply FISTA. In particular, we employ the composite splitting approach in \cite{fista_l1_tv} to decompose the above problem into two sub problems. The first problem is
\begin{equation}\label{eq17}
\mathbf{\hat{x}}_1^{\prime}=\arg\min_{\mathbf{x}^{\prime}}\left \{ \frac{w_1 L}{2}\Vert {\mathbf{x^{\prime}}-\mathbf{v}_k}\Vert_2^2+{\lambda_1}\Vert{\mathbf{x'}}\Vert_1 \right \}, 
\end{equation}
and the second is 
\begin{equation}\label{eq18}
\mathbf{\hat{x}}_2^{\prime}=\arg\min_{\mathbf{x}^{\prime}}\left \{ \frac{w_2 L}{2}\Vert {\mathbf{x^{\prime}}-\mathbf{v}_k}\Vert_2^2+\lambda_{\textrm{TV}}\Vert{\mathbf{x'}}\Vert_{\rm{TV}} \right \}, 
\end{equation}
where $w_1$ and $w_2$ are the decomposition weights that separate $\frac{L}{2}\Vert {\mathbf{x^{\prime}}-\mathbf{v}_k}\Vert_2^2$ into two groups associated with the $\ell_1$ norm and the TV norm, respectively, and satisfying $w_1+w_2=1$. 

Equations \eqref{eq17} and \eqref{eq18} can be readily solved by the algorithms in \cite{fista_tv}. The final solution is $\mathbf{\hat{x}}^{\prime}=w_1 \mathbf{\hat{x}}_1^{\prime} + w_2 \mathbf{\hat{x}}_2^{\prime}$. 
The different steps in the proposed MRF based FISTA algorithm are given in $\textbf{Algorithm 1}$. Here, we assume $w_1\!=\!w_2\!=\!1/2$.
\renewcommand{\algorithmicrequire}{\textbf{Input:}}
\renewcommand{\algorithmicensure}{\textbf{Output:}}
\begin{algorithm}[htb]
	\caption{ : MRF-FISTA } 
	\label{alg:Framwork} 
	\begin{algorithmic} 
		\REQUIRE 
		$L$-An upper bound on the Lipschitz constant of $\nabla f$;\\
		\ENSURE 
		Recovered signal: $\mathbf{\hat{x}}=\mathbf{x}_k$;
		\renewcommand{\algorithmicrequire}{\textbf{Step 0:}}
		\REQUIRE 
		Take $\mathbf{z}_1=0$, $t_1=1$, $w_1=w_2=1/2$.
		\renewcommand{\algorithmicrequire}{\textbf{Step k:}}
        \REQUIRE 	
        Compute ($k\geq 1$)
		\STATE 
$\mathbf{x}_{1,k}=\arg\min\limits_{\mathbf{x}}\Big\{\lambda_1\|\mathbf{x}\|_1+\frac{w_1 L}{2}\|\mathbf{x}-(\mathbf{z}_k-\frac{1}{L}\nabla f)\|_2^2 \Big\}$
		\STATE 
$\mathbf{x}_{2,k}=\arg\min\limits_{\mathbf{x}}\Big\{\lambda_{\textrm{TV}}\|\mathbf{x}\|_{\textrm{TV}}+\frac{w_2 L}{2}\|\mathbf{x}-(\mathbf{z}_k-\frac{1}{L}\nabla f)\|_2^2 \Big\}$		 
		\STATE $\mathbf{x}_k^{\prime}=(\mathbf{x}_{1,k}+\mathbf{x}_{2,k})/2$ 
	
        \STATE $\mathbf{s}_k= \textrm{MAP - support}\{\mathbf{x}_k^{\prime}\}$
        
		\STATE $\mathbf{x}_k^{\prime}=\mathbf{x}_k^{\prime} \circ \mathbf{s}_k$

		\STATE $\mathbf{x}_k=\arg\min\limits_{\mathbf{x}}\{F(\mathbf{x}): \mathbf{x}=\mathbf{x}_k^{\prime}, \mathbf{x}_{k-1}\}$, where $F(\mathbf{x})=\frac{1}{2}\Vert {\mathbf{y}\!\!-\!\!\mathbf{\Phi x}}\Vert_2^2\!\!+\!\!{\lambda_1}\Vert{\mathbf{x}}\Vert_1\!\!+\!\!\lambda_{\textrm{TV}}\Vert{\mathbf{x}}\Vert_{\rm{TV}}$

		\STATE $t_{k+1}=\frac{1+\sqrt{1+4t_k^2}}{2} $
        
		\STATE $\mathbf{z}_{k+1}=\mathbf{x}_k+\frac{t_k}{t_{k+1}}(\mathbf{x}_k^{\prime}-\mathbf{x}_k)+\frac{t_k-1}{t_{k+1}}(\mathbf{x}_k-\mathbf{x}_{k-1})$

	\end{algorithmic}
\end{algorithm}

\section{MRF-FISTA Applied to ISAR Imaging}

\begin{figure}[!t]
	\centering
	\includegraphics[width=1.7in]{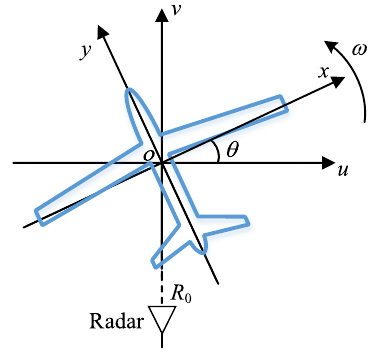}
	\caption{ISAR imaging geometry.}
\label{fig3}
\end{figure}

The ISAR geometry, after compensating for the translational motion, can be shown in Fig. \ref{fig3} in which the target rotates around its centroid with a constant angular velocity $\omega$. 
The scattered electromagnetic (EM) wave after demodulation is,
\begin{equation}\label{radar_sig}
Y(f_n,t_m)=\sum_{i=1}^I \sigma_i\exp\left[-\mathrm{j2\pi}f_n\frac{2R_i(t_m)}{c}\right],
\end{equation}
where $\sigma_i$ denotes the backscattering coefficient of the $i\mathrm{th}$ point scatterer, $I$ is the number of total pixels of the target scene, $f_n$ represents the operating frequency: $f_n\!=\!f_0\!+\!n\Delta f$, $f_0$ and $\Delta f$ denote the start frequency and the frequency step of the transmitted signal, respectively, 
and $n$ is an integer from the set $[0,N-1]$. The variable  $t_m$ represents the discrete slow time in the cross-range dimension, with $m$ is an integer from the set $[0,M-1]$, $c$ is the speed of EM wave propagating in free space, and $R_i(t_m)$ is given by,
\begin{equation}\label{eq20}
R_i(t_m)\!=\!\sqrt{R_0^2+x_i^2+y_i^2+2R_0[-x_i\sin {\omega t_m}+y_i\cos{\omega t_m}]},
\end{equation}
representing the instantaneous range between the $i\mathrm{th}$ scatterer and the radar, where $R_0$ denotes the distance from the target centroid to the radar.

According to \eqref{radar_sig}, we can formulate the element of the sensing matrix $\mathbf{\Phi}$ as, $\phi_{l,i}=\exp\big [\mathrm{-j2\pi}f_n\frac{2R_i(t_m)}{c} \big]$, where $n=(l \mod N)$, and $m=\lfloor \frac{l}{M} \rfloor$. Here, ``mod" means the modulo operation, and $``\lfloor \rfloor"$ denotes rounding toward negative infinity, and $l=0,1,\cdots,MN-1$, $i=0,1,\cdots,I-1$. 

The target scene can be viewed as a matrix $\mathbf{X}$ with a size of $N\times M$ (assuming $I=N\times M$). Thus, the forward model of radar imaging can be expressed as, $\textrm{vec}(\mathbf{Y})=\mathbf{\Phi} \textrm{vec}(\mathbf{X})$, where $\mathbf{Y}$ denotes the matrix form of the received data (with zeros padding at the positions where there are no samples), and $\textrm{vec}(\cdot)$ represents vectorization of a matrix. Then, the image $\mathbf{X}$ can be reconstructed by the proposed MRF-FISTA. 

If the size of the sensing matrix is too large, we can use the method in \cite{li1} to change $\mathbf{\Phi}$ into a 2-D Fourier transform with a compensation for the spherical wavefront. In addition, if the distance $R_0$ is sufficiently large, \eqref{eq20} can be approximated as $R_i(t_m)\!\approx \!R_0\!-\!x_i\sin {\omega t_m}\!+\!y_i\cos{\omega t_m}$. 
In this case, the sensing matrix can also be replaced by a  Fourier transform.

\section{Simulations and Experimental Results}
This section shows the performance of the MRF-FISTA based ISAR imaging method using both simulation and real data.

\subsection{Simulation results}
\begin{figure}[!t]
	\centering
	\includegraphics[width=2.5in]{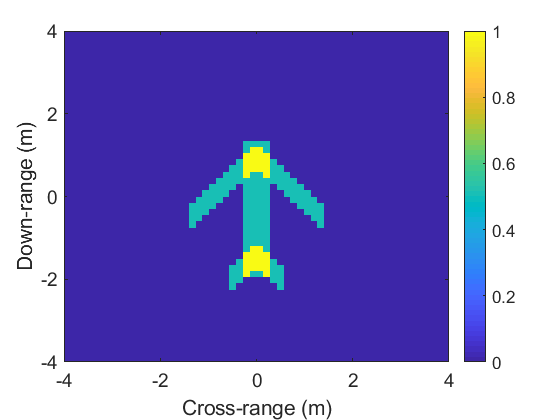}
	\caption{Target model.}
\label{fig5}
\end{figure}

The simulation parameters for the full data set, according 
to the Nyquist-Shannon sampling theorem, are shown in Table \ref{tab1}.
The target model is shown in Fig. \ref{fig5}.

\begin{table}
\centering
\caption{Simulation Parameters for Full Data Set}
\label{tab1}
\setlength{\tabcolsep}{3pt}
\begin{tabular}{|p{90pt}|p{40pt}|}
\hline
Parameters& 
Values \\
\hline
Distance $R_0$&
5 km\\
Start frequency& 
35 GHz \\
Stop frequency&
36 GHz \\
Number of frequency steps&
64 \\
Total rotation angle&
$1.7^{\circ}$ \\
Number of angle steps&
64\\

\hline
\end{tabular}
\end{table}

\begin{figure}[!t]
	\centering
    \subfloat[]{
	\includegraphics[width=1.7in]{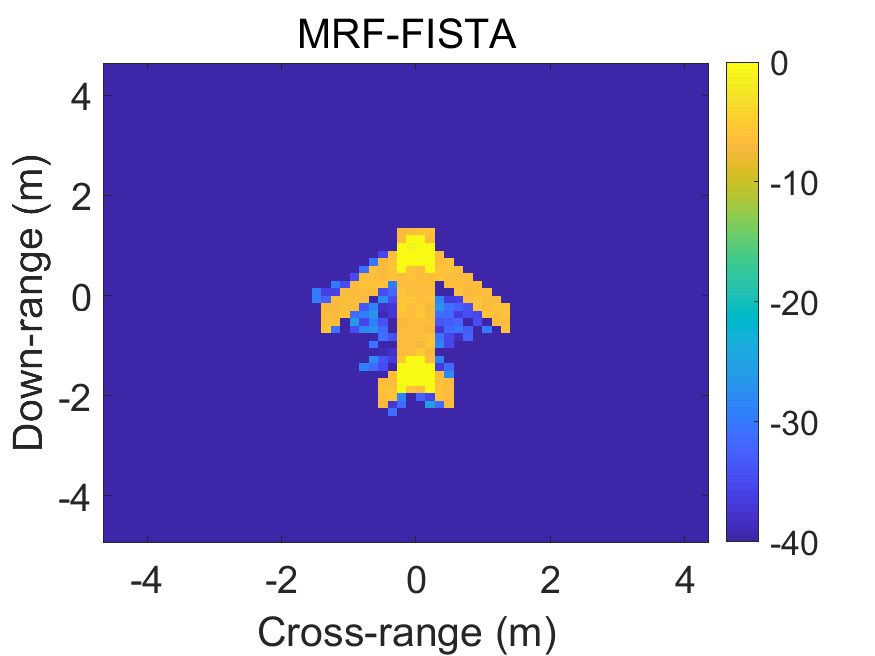}}
        \label{5a}\hfill
  \subfloat[]{%
        \includegraphics[width=1.7in]{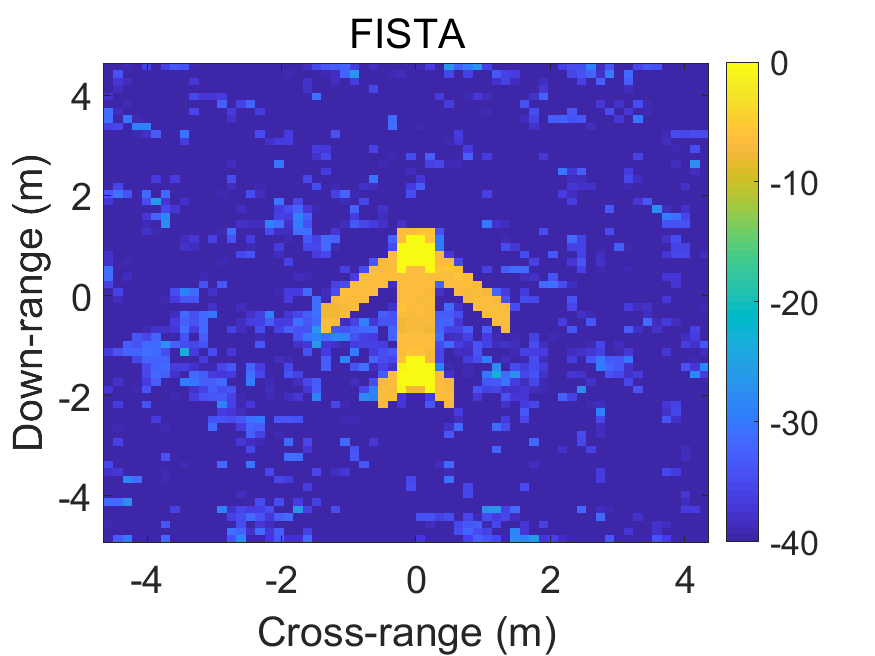}}
    \label{5b}\\
    \subfloat[]{
	\includegraphics[width=1.7in]{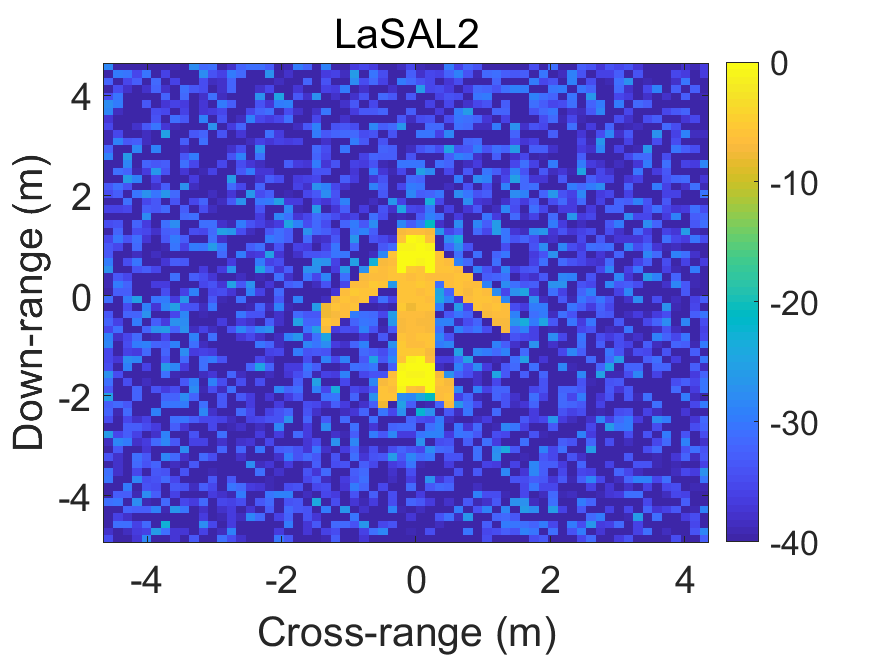}}
        \label{5c}\hfill
  \subfloat[]{%
        \includegraphics[width=1.7in]{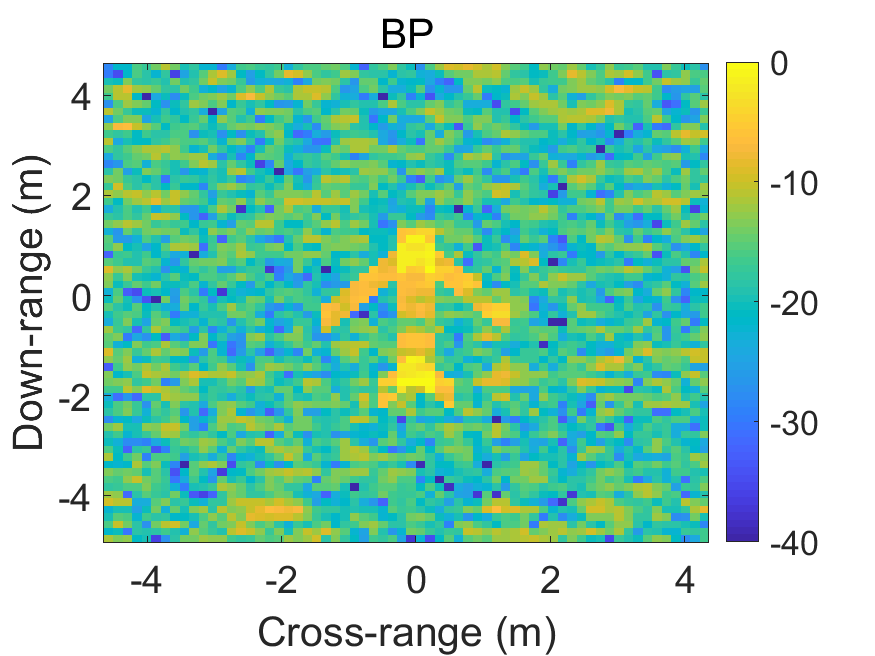}}
    \label{5d}\\
	\caption{Imaging results by (a) MRF-FISTA, (b) FISTA, (c) LaSAL2, and (d) BP, respectively, using 30\% of the full data with SNR=20dB.}
\label{fig6}
\end{figure}

First, we present the imaging results of four techniques, two of which use MRF prior. We consider the proposed MRF-FISTA, FISTA,  LaSAL2, which is a recently introduced ADMM based algorithm with the MRF prior\cite{b30}, and the traditional back-projection (BP) algorithm, respectively, in Fig. \ref{fig6}.
The sampling rate (i.e., the ratio of the utilized data size over the full data set size) is 30\%. The signal-to-noise ratio (SNR) is 20 dB. Note that the imaging result of MRF-FISTA is vividly of higher quality than those of the other three algorithms.

In order to show the performance sensitivity of the proposed method and LaSAL2 with respect to their corresponding penalty parameters, we compare the convergence of the algorithms with different penalties.
Here, we employ the root mean squared error (RMSE) to assess convergence, 
which is given by, $\textrm{RMSE}\!\!=\!\!\sqrt{\frac{1}{NM}\sum_{n=0}^{N-1}\sum_{m=0}^{M-1} \big [ \mathbf{X}(n,m) \!\!-\!\! \mathbf{\hat{X}}(n,m) \big ]^2}$, 
where $\mathbf{X}$ and $\mathbf{\hat{X}}$ denote the target model and the recovered image, respectively. 

\begin{figure}[!t]
	\centering
    \subfloat[]{
	\includegraphics[width=1.7in]{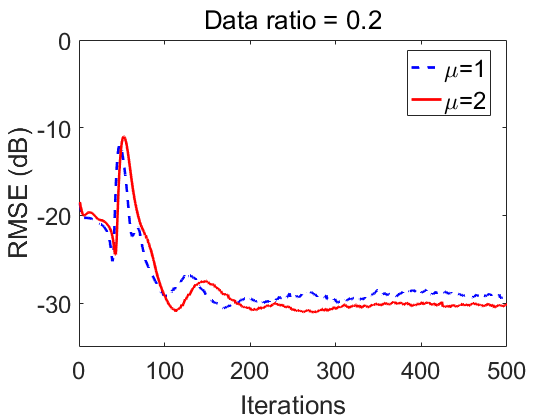}}
        \label{8a}\hfill
  \subfloat[]{%
        \includegraphics[width=1.7in]{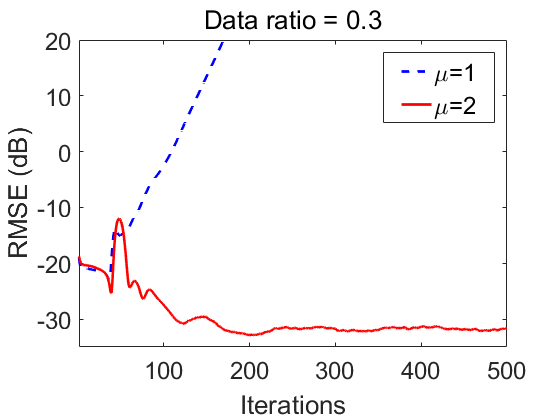}}
    \label{8b}\\

	\caption{RMSEs of LaSAL2 with respect to different penalty parameters $\mu$ by using (a) 20\%, (b) 30\% of the full data set.}
\label{fig_admm_convergence}
\end{figure}

\begin{figure}[!t]
	\centering
    \subfloat[]{
	\includegraphics[width=1.7in]{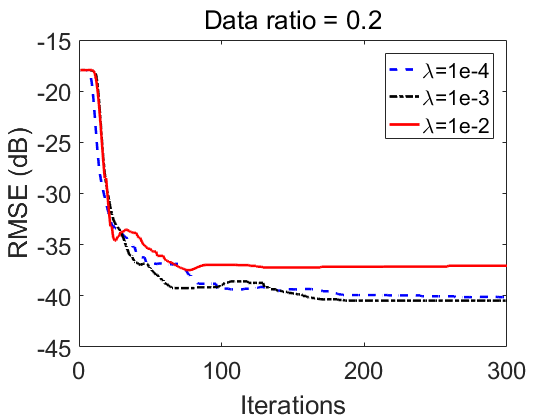}}
        \label{9a}\hfill
  \subfloat[]{%
        \includegraphics[width=1.7in]{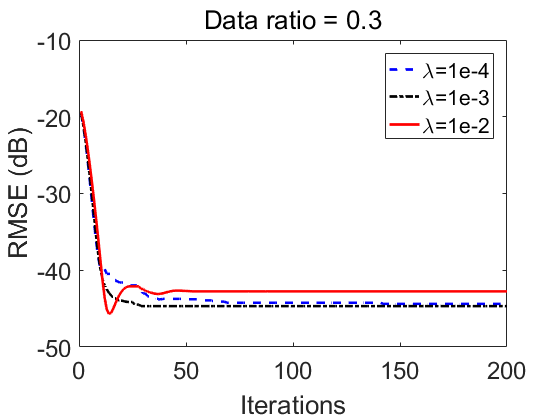}}
    \label{9b}\\

	\caption{RMSEs of MRF-FISTA with respect to different penalty parameters $\lambda$ by using (a) 20\%, (b) 30\% of the full data set.}
\label{fig_MRF-FISTA_convergence}
\end{figure}

Fig. \ref{fig_admm_convergence} shows the RMSEs of LaSAL2 with different starting value of the penalty parameters, $\mu$, which underlies the augmented Lagrangian, for different data sets. 
The SNR is 20 dB for this simulation. Note that smaller penalty parameters can result in divergence for larger data sets. A larger penalty has a better solution, however, with a slower convergence rate. Thus, for such methods, it becomes critical to choose a proper starting penalty parameter. 

\textcolor{black}{Unlike the ADMM based algorithms, the penalty parameter $\lambda$ of FISTA is only employed to tradeoff the $l_2$ error and the signal sparsity.
}
Fig. \ref{fig_MRF-FISTA_convergence} shows the RMSEs of MRF-FISTA with different $\lambda$s. 
Clearly, the convergence rate almost remains invariant with different penalty parameters. 

Next, we include another performance metric -- \textit{entropy}, for the quantitative comparisons.
The Shannon entropy of an image $\mathbf{X}$ is defined as, $H(\mathbf{X})=-\sum_{k=1}^K p_k(\mathbf{X}_k) \log_2 p_k(\mathbf{X}_k)$, 
where $(p_1,p_2,\cdots,p_K)$ is a finite discrete probability distribution, and $p_k$ contains the normalized histogram counts within a fixed bin $\mathbf{X}_k$, i.e., it is a function of pixel intensity in an image. The distribution of pixel intensity is commensurate with the degree of image focus. 

\begin{figure}[!t]
	\centering
    \subfloat[]{\label{fig7a}
	\includegraphics[width=1.69in]{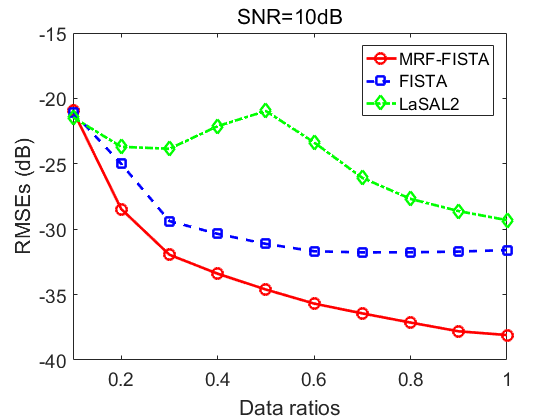}}
        \hfill
  \subfloat[]{\label{fig7b}
        \includegraphics[width=1.69in]{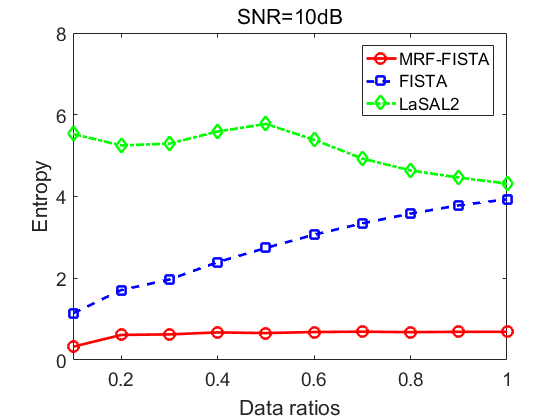}}
    \\
	\caption{(a) RMSE, and (b) entropy of the reconstructed images versus different data ratios.}
\label{fig8}
\end{figure}

Fig. \ref{fig8} shows the RMSEs and entropy of the imaging results obtained by the MRF-FISTA, FISTA, and the LaSAL2, respectively, for different data ratios. The SNR is fixed to 10 dB. We perform 50 independent runs for each data ratio to obtain the  mean values of the RMSEs and entropy. It is noted from Fig. \ref{fig8}\subref{fig7a} that the MRF-FISTA improves clearly over FISTA and LaSAL2 more than at least 3 dB and 4 dB, respectively, when the data ratios are above 20\%.   
Also, the entropy of the results obtained by the MRF-FISTA is much smaller than those of FISTA and LaSAL2, indicating that the imaging results of MRF-FISTA are best focused among these three algorithms, as demonstrated in Fig. \ref{fig8}\subref{fig7b}.

The parameters of the Ising model were chosen as: $\alpha=0.01$ and $\beta=0.3$ for the above simulations.

\subsection{Experimental results}
Here, we use measurements from an unmanned aerial vehicle. The radar frequencies in the experiments vary from 34.5 GHz to 35.3 GHz with 128 steps. The total  observation angles are 2 degrees, also with 128 steps. 

\begin{figure}[!t]
	\centering
    \subfloat[]{
	\includegraphics[width=1.7in]{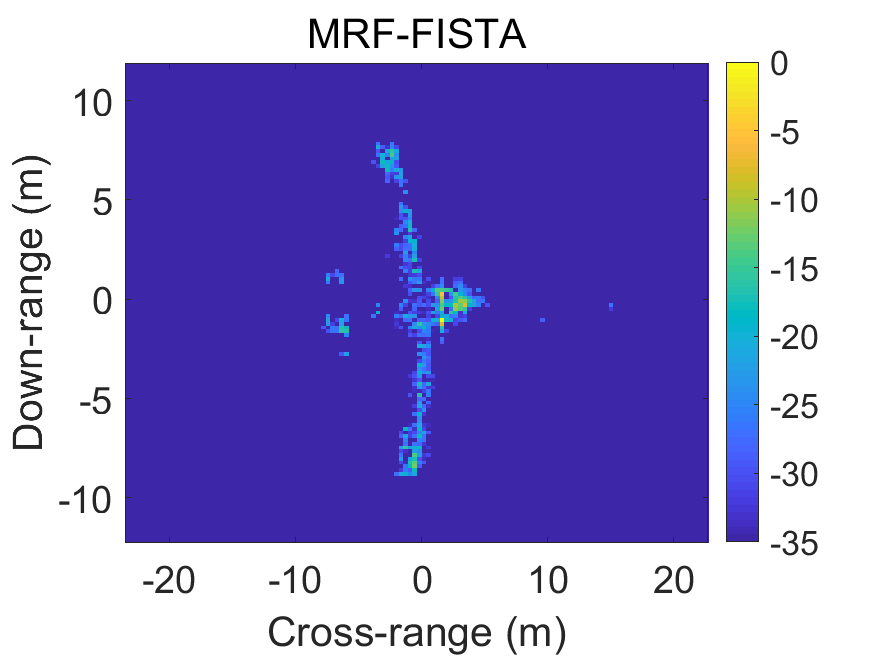}}
        \label{11a}\hfill
  \subfloat[]{%
        \includegraphics[width=1.7in]{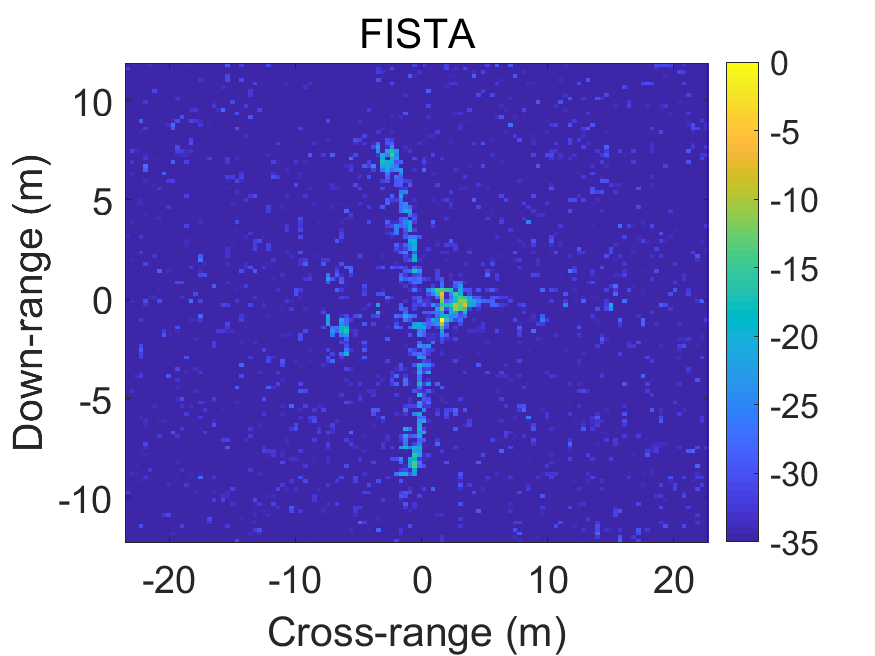}}
    \label{11b}\\
    \subfloat[]{
	\includegraphics[width=1.7in]{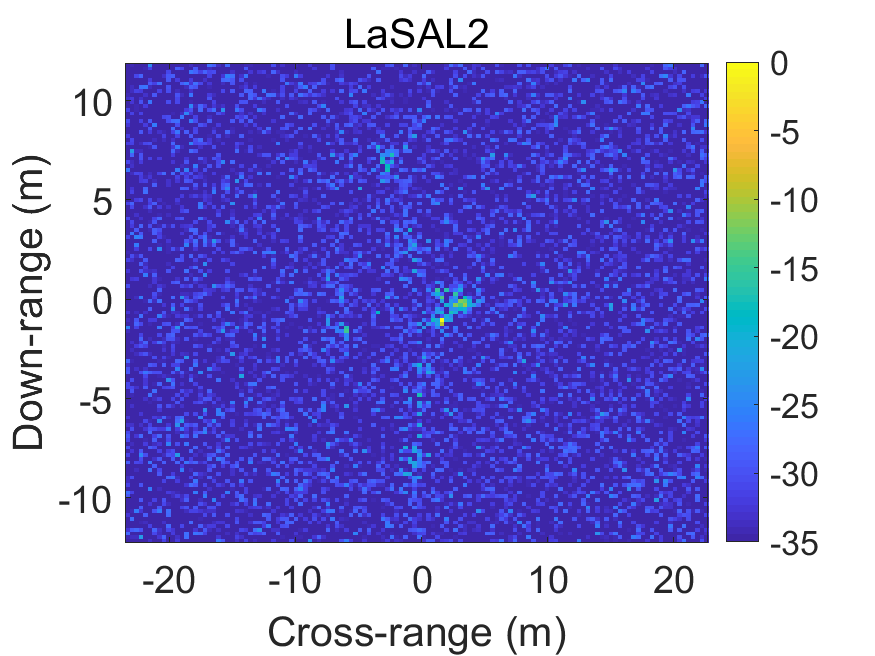}}
        \label{11c}\hfill
  \subfloat[]{%
        \includegraphics[width=1.7in]{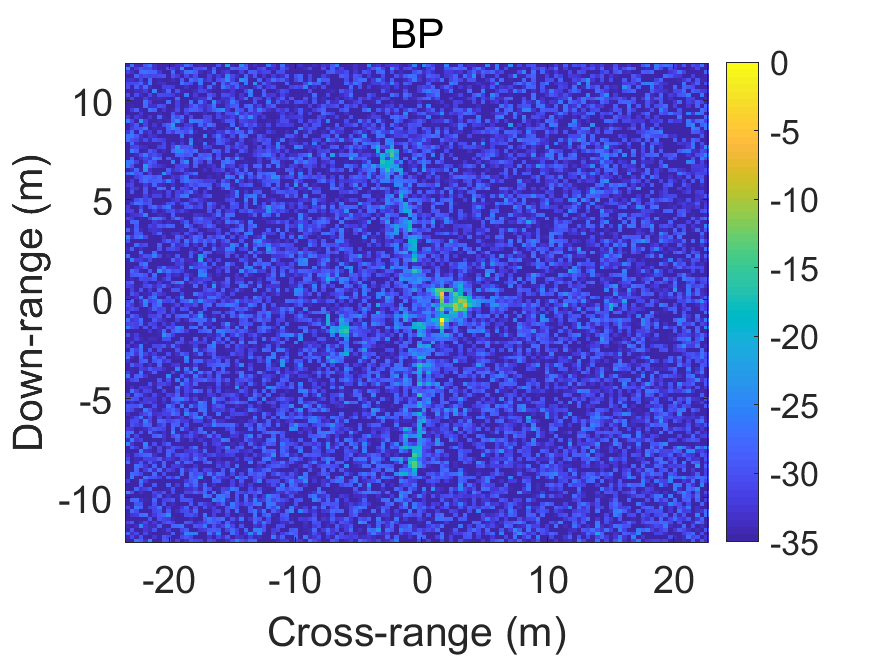}}
    \label{11d}\\
	\caption{Imaging results by (a) MRF-FISTA, (b) FISTA, (c) LaSAL2, and (d) BP, respectively, using 30\% of the full data. }
\label{fig12}
\end{figure}

\begin{figure}[!t]

  \centering
        \includegraphics[width=2.5in]{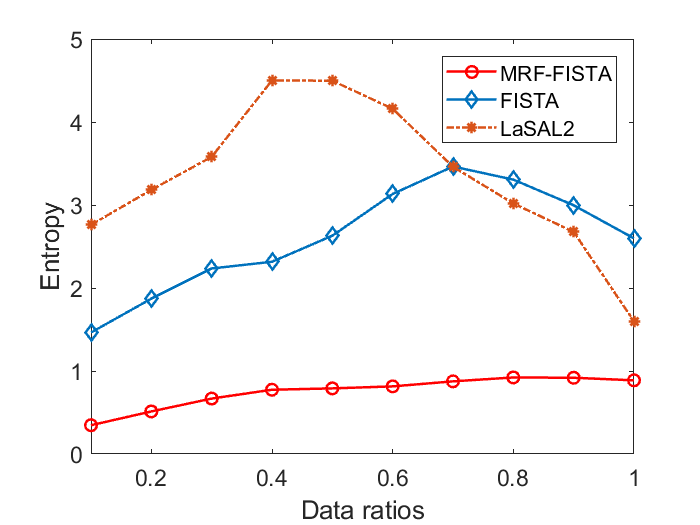}
    \label{13b}\\
	\caption{Entropy of the reconstructed images versus different data ratios.}
\label{fig14}
\end{figure}

Fig. \ref{fig12} depicts the imaging results of MRF-FISTA, FISTA,  LaSAL2 and BP, using 30\% of the full sampled data. It is seen that the result of MRF-FISTA assumes the best quality with much smaller noisy like clutter among all the algorithms. 

Since we do not have a bench mark for the real data imaging, we present the entropy results of the images to assess the quantitative performance of the algorithms when processing the real data. This is shown in Fig. \ref{fig14}. 
Clearly, the precision of MRF-FISTA is still higher than those of FISTA and LaSAL2, with much smaller entropy.

\section{Conclusion}
This letter proposed a MRF based FISTA as an effective image reconstruction technique for cluster-type images, including those of ISAR. The MRF is a powerful tool for modeling spatial context dependent entities. It was used in this letter as a prior to describe the high spatial correlations in an image.
The MRF prior was incorporated into sparsity based optimization algorithm, and shown to yield higher image reconstruction precision compared to existing techniques. 
Simulations and experimental results demonstrated that the proposed algorithm improves over the original  FISTA and another recently introduced MRF based ADMM algorithm when considering imaging errors as well as convergence rate. 


\bibliographystyle{full} 
\bibliography{full}

\end{document}